\documentclass[english]{paper}
\usepackage[T1]{fontenc}
\usepackage[latin9]{inputenc}
\usepackage{babel}

\usepackage{mathrsfs}
\usepackage{url}
\usepackage{amstext}
\usepackage{esint}
\usepackage[unicode=true, pdfusetitle,
 bookmarks=true,bookmarksnumbered=false,bookmarksopen=false,
 breaklinks=false,pdfborder={0 0 1},backref=false,colorlinks=false]
 {hyperref}
\begin{document}

\title{Non-holonomic Constraint Force Postulates}

\author{Bharath H M, Indian Institute Of Technology, Kanpur}
\maketitle
\begin{abstract}
The extended Hamilton's Priciple and other methods proposed to handle
non-holonomic constraints are considered. They dont agree with each
other. By looking at its consistency with D'Alembert principle for
linear non-holonomic constraints, it was claimed in earlier papers
that the direct extension of hamilton's principle is incorrect. Nonholonomic
Constraints, linear in velocities were considered for this purpose.
This paper analyzes these claims, and shows that they are incorrect.
And hence it shows that it is theoretically impossible to judge any
attempt on non-holonomic constraints to be wrong, as long as they
are consistent with the D'Alembertian for holonomic constraints.
\end{abstract}

\section*{I. Introduction}

The problem of mechanics with non-holonomic constraints was considered
recently in \cite{key-3,key-2} and \cite{key-5}. \cite{key-1} and
\cite{key-9}, appear to be among the earliest to propose a direct
extension of the hamilton's principle to the non-holonomic regime.
The same extension was adopted in \cite{key-12}. However, later papers
(\cite{key-3,key-2,key-8} to name a few) attempted to show that such
an extension is incorrect. \cite{key-3} appears to be very conclusive
regarding this issue. It has ruled out the extension of hamilton's
principle usimg a theoritical argument. The present paper shows that
such theoritical arguments cannot rule out the extension.\medskip{}

This paper discusses the extension of hamilton's principle along with
other attempts to tackle non-holonomic constraints, and carefully
examines the argument used to rule out the direct use of hamilton's
principle for non-holonomic constraints. Finally, it points out the
reason why the argument is incorrect. That slightly alters the conclusions
made in \cite{key-3} regarding the scope of the D'Alembert's principle.
These conclusions show that the two ways of tackling non-holonomic
problems are uncomparable.\medskip{}

The paper first briefly discusses the classical theory of mechanics,
which can be found in any classical text on mechanics\cite{key-12}.
However, the following section-II outlines the theory in a structure
convenient for the present purpose. Section III analyzes the various
proposals to tackle non-holonomic constraints. Section IV analyzes
the argument given against a direct generalization of the Hamilton's
principle, and points out the fallacy in it. Section V summarizes
its implications. The mathematical tools required, namely, the variational
calculus is put at the end, in the appendix. Einstein's notation is
used for summation throughout. Unless explicitly mentioned, repeated
indices mean summation throughout. \medskip{}

\section*{II. Holonomic Constraints\medskip{}
}

\subsection*{1. Newton's Laws}

For a system of particles Newton's law can be written under a Cartesian
system as: (indices not summed)

\[
m_{i}\ddot{x_{i}}=X_{i}\]
Where $X_{i}$ are the total forces. Newton's law says, acceleration
is caused by force and only force. Hence, within the framework of
newtons laws, nothing other than a force can influence the dynamics
of a system. Any constraint which restricts the motion of a system
should hence be replaced by an equivalent 'constraint force' which
brings in the same effect as the constraint, so that it can be studied
using Newton's laws. The equation becomes $m_{i}\ddot{x_{i}}=F_{i}+Q_{i}$,
where $F_{i}$ are the applied forces and$Q_{i}$ are the constraint
forces. An additional postulate is required to construct the appropriate
constraint forces. 

\medskip{}

\subsection*{2. Constraint Force Postulate (D'Alembert's Principle): }

For a constraint of the form $\text{\ensuremath{\mathscr{F}}}(x,t)=0$,
which means, that the system is confined to a time varying hyper surface
in the configuration space, the constraint forces $Q_{i}$ which have
the same effect as that of the constraint can be written as,\begin{equation}
Q_{i}=\lambda(t)\frac{\partial}{\partial x_{i}}\text{\ensuremath{\mathscr{F}}}\label{A}\end{equation}
 where $\lambda(t)$ is an undetermined function of time. This formulation
of the postulate will be shown to be equivalent to the standard D'Alembert's
principle stated in terms of virtual work, defined in most standard
texts on mechanics( \cite{key-12}, for example). 

\medskip{}

A 'virtual displacement' is a local variation of $x(t)$ consistent
with the constraint. i.e., a local variation $\{\delta_{i}(t)\}$
such that, $\text{\ensuremath{\mathscr{F}}}_{x_{i}}\delta_{i}=0$
,assuming that not all $\text{\ensuremath{\mathscr{F}}}_{x_{i}}$are
zero, at a time t. This is a simpler form of the definition given
by \cite{key-16}. It says {}``virtual displacement is a vector tangential
to the constraint manifold'' The definition of a virtual displacement
is found all texts on mechanics. However, most of them are unclear
and ambiguous. Hence we refer to \cite{key-16}. As it turns out,
the major result pointed out by this paper banks on this definition,
hence this extra care on the definition. D'Alembert's principle says
the virtual work done by the constraint forces vanishes\cite{key-12}.
i.e., $Q_{i}\delta_{i}=0,$ at a time t for all $\{\delta_{i}(t)\}$
such that $\text{\ensuremath{\mathscr{F}}}_{x_{i}}\delta_{i}=0$ Or,
$Q_{i}=\lambda\text{\ensuremath{\mathscr{F}}}_{x_{i}}$ where the
constant $\lambda$ might not be the same at a different time t'.
This can be in general written as $Q_{i}=\lambda(t)\text{\ensuremath{\mathscr{F}}}_{x_{i}}$.
Hence equation \ref{A} is equivalent to D'Alembert's principle. The
equations of motion can now be written as \begin{equation}
m_{i}\ddot{x_{i}}=F_{i}+\lambda(t)\text{\ensuremath{\mathscr{F}}}_{x_{i}}\label{1}\end{equation}
 where $F_{i}$ are the applied forces. 

\medskip{}

\subsection*{3. Variational Formulation}

For a system where the applied forces are conservative, the equation
of motion \ref{1} supports a variational formulation. If \ensuremath{\mathscr{L}}
is the Lagrangian, the equations \ref{1} can be written as (\cite{key-12}): 

\begin{equation}
\frac{d}{dt}\text{\ensuremath{\mathscr{L}}}_{\dot{x_{i}}}-\text{\ensuremath{\mathscr{L}}}_{x_{i}}=\lambda(t)\text{\ensuremath{\mathscr{F}}}_{x_{i}})\label{2}\end{equation}
Since \ensuremath{\mathscr{F}} is independent of the velocities, $\lambda(t)\text{\ensuremath{\mathscr{F}}}_{x_{i}}=-\{\frac{d}{dt}\frac{\partial}{\partial\dot{x_{i}}}(\lambda(t)\text{\ensuremath{\mathscr{F}}})-\frac{\partial}{\partial x_{i}}(\lambda(t)\text{\ensuremath{\mathscr{F}}})\}$
. Hence the equations can be written as $\frac{d}{dt}\frac{\text{\ensuremath{\partial}}}{\text{\ensuremath{\partial}}\dot{x_{i}}}(\text{\ensuremath{\mathscr{L}}}+\lambda(t)\text{\ensuremath{\mathscr{F}}})-\frac{\text{\ensuremath{\partial}}}{\text{\ensuremath{\partial}}x_{i}}(\text{\ensuremath{\mathscr{L}}}+\lambda(t)\text{\ensuremath{\mathscr{F}}})=0$.
Which can be compared with equation \ref{eq:17} in the appendix.
These equations suggest the Hamilton's principle of stationary action
for this category of constraints. The equation \ref{eq:17} along
with initial and final condition (i.e, the \textit{boundary value}
problem) is an extremization of the action. However, the corresponding\textit{
initial value} problem (i.e, the same differential equation with initial
position and velocities known, which is the case in all physical situations)
is not strictly equivalent to extremization of the action. However,
the Hamilton's principle can be formulated with the obtained boundary
values as 'action is stationary for fixed boundaries'. \medskip{}

\section*{III. Non Holonomic Constraints}

The D'Alemberts virtual work postulate stated in the previous section
fails to work with nonholonomic constraints. A new constraint force
postulate is required to handle non holonomic constraints. Various
postulates are available in literature\cite{key-14}. The major candidates
are:\medskip{}

\subsection*{1.) Gauss-Gibbs Principle}

The Gauss-Gibbs principle as described in \cite{key-14,key-6}, says
that the first order variation of the quantity, C, vanishes under
allowed Gaussian variations. \begin{equation}
C=\frac{1}{2}m_{i}(\ddot{x_{i}}-\frac{X_{i}}{m_{i}})^{2}\label{3}\end{equation}
Variations $\{\delta_{i}(t)\}$ for which $\delta_{i}$ and $\dot{\delta_{i}}$
vanish at the concerned value of t are called Gaussian variations.
Under such variations, the first order variation of C is given by:

\begin{equation}
\delta C=(m_{i}\ddot{x_{i}}-X_{i})\ddot{\delta_{i}}\label{4}\end{equation}
This quantity vanishes for all Gaussian variations consistent with
constraints. The consistency condition translates as:\begin{equation}
\text{\ensuremath{\mathscr{F}}}(x+\delta,\dot{x}+\dot{\delta},t)-\text{\ensuremath{\mathscr{F}}}(x,\dot{x},t)=\text{\ensuremath{\mathscr{F}}}_{x_{i}}\delta_{i}+\text{\ensuremath{\mathscr{F}}}_{\dot{x_{i}}}\dot{\delta_{i}}=0\label{5}\end{equation}
Which is trivially true at the concerned time t. And hence we turn
to its first order variation in time.\begin{equation}
\delta\text{\ensuremath{\mathscr{F}}}(t+\delta t)-\delta\text{\ensuremath{\mathscr{F}}}(t)=\{\text{\ensuremath{\mathscr{F}}}_{\dot{x_{i}}}\ddot{\delta_{i}}+(\dot{\text{\ensuremath{\mathscr{F}}}_{\dot{x_{i}}}}+\text{\ensuremath{\mathscr{F}}}_{x_{i}})\dot{\delta_{i}}+\dot{\text{\ensuremath{\mathscr{F}}}_{x_{i}}}\delta_{i}\}\delta t=0\label{eq:6}\end{equation}
That puts the condition $\text{\ensuremath{\mathscr{F}}}_{\dot{x_{i}}}\ddot{\delta_{i}}=0$
on the $\ddot{\delta_{i}}$ (Second order time variation needs to
be invoked for holonomic constraints). The principle is now equivalent
to newton's laws with a force postulate: \begin{equation}
Q_{i}=\lambda(t)\text{\ensuremath{\text{\ensuremath{\mathscr{F}}}}}_{\dot{x_{i}}}..........(nonholonomic)\label{eq:B1}\end{equation}
 \begin{equation}
Q_{i}=\lambda(t)\text{\ensuremath{\text{\ensuremath{\mathscr{F}}}}}_{x_{i}}.............(holonomic)\label{eq:B2}\end{equation}
 A similar result is obtained in \cite{key-4}.\medskip{}

\subsection*{2.) Jordain Principle}

The fundamental equation of the Jourdain principle is(\cite{key-7,key-14})\begin{equation}
(m_{i}\ddot{x_{i}}-X_{i})\dot{\delta_{i}}=0\label{eq:7}\end{equation}
Where $\delta_{i}(t)$ are a 'Jourdain' variations which vanish at
t, while their derivatives dont. Such variations are called 'virtual
velocities'. The constraint force postulate used is same as those
in the Gauss's principle(\ref{eq:B1},\ref{eq:B2})\medskip{}

\subsection*{3.) Extended Hamilton's principle: Vakonomic Mechanics}

Extending Hamilton's principle of stationary action was proposed in
\cite{key-1,key-9}, later criticized heavily (\cite{key-2,key-3,key-8}
for example.) Nevertheless, it is equivalent to assuming a constraint
force given by \begin{equation}
Q_{i}=\lambda(t)\{\text{\ensuremath{\mathscr{F}}}_{x_{i}}-\frac{d}{dt}\text{\ensuremath{\mathscr{F}}}_{\dot{x_{i}}}\}+\text{\ensuremath{\mathscr{F}}}_{\dot{x_{i}}}\frac{d\lambda(t)}{dt}\label{eq:C}\end{equation}
This was termed as 'vakonomic mechanics'(\textbf{V}ariational\textbf{
A}xiomatic \textbf{K}ind) by \cite{key-17} since the basic axiom
here is the Hamilton's principle, which is variational in nature.

We now have two possible postulations of the the constraint force:
\ref{eq:B1}\ref{eq:B2} and \ref{eq:C}. We have one well accepted
formulation \ref{A}(D'Alembert's principle) for holonomic constraints.
Both \ref{eq:B1}\ref{eq:B2} and \ref{eq:C} agree with \ref{A}
for the special case of holonomic constraints. But, they don't at
proper nonholonomic constraints. One curious and simple set of constraints
which might be of help in deciding the correctness of \ref{eq:B1}\ref{eq:B2}
and \ref{eq:C} is the linear nonholonomic constraints. This has been
used in \cite{key-2,key-3} to rule out \ref{eq:C}. However, there
seems to be a subtle issue with that and hence we wish to reconsider
it.\medskip{}

\section*{IV. Linear Non-Holonomic Constraints}

Linear nonholonomic constraints are non integrable equations of the
form,\begin{equation}
a_{i}\dot{x_{i}}+a_{t}=0\label{eq:8}\end{equation}
Where $a_{i}$ and $a_{t}$ are functions independent of $\dot{x}_{i}$.
This category of constraints has been recently considered in \cite{key-3}
to discard \ref{eq:C}. \cite{key-3} has concluded that all nonholonomic
constraints are beyond the scope of Hamilton's principle; the general
nonholonomic constraints are beyond the scope of D'Alemberts principle,
while the linear non holonomic constraints do remain within the scope
of D'Alembert's principle. And they agree with what \ref{eq:B1}\ref{eq:B2}
says; and \ref{eq:C} is inconsistent with it, hence it is incorrect.
This has been done in other literature as well. \cite{key-3} examines
the reason behind it's conclusions. And it says, the reason why \ref{eq:C}
fails at a linear non-holonomic is that the \textit{allowed} variations
are \textit{not consistent} with the constraints. By allowed variations,
we mean the variations constrained by \begin{equation}
a_{i}\delta_{i}=0\label{eq:9}\end{equation}
 which is indeed obtained from the constraint equation, but not directly
stating that the variations are consistent with the constraint. But,
the constraint force given by \ref{eq:9} is $Q_{i}=\lambda(t)a_{i}$.
Which is same as what is obtained from \ref{eq:B1}\ref{eq:B2}, as
mentioned above. However, \ref{eq:B1}\ref{eq:B2} obtaines the constraint
force through a different route, not from \ref{eq:9}. Since equation
\ref{eq:9} is mysterious, we take a look at how to arrive at \ref{eq:9}
from \ref{eq:8}.

It is widely mentioned that \ref{eq:9} is obtained from \ref{eq:8}
using D'Alembert's principle of virtual work(\cite{key-2,key-3,key-10,key-13,key-14,key-15}).
They write \ref{eq:8} as\begin{equation}
a_{i}dx_{i}+a_{t}dt=0\label{eq:10}\end{equation}
And then replace $dx_{i}$ by $\delta_{i}$ and $dt$ by 0. But the
trouble is, this cannot be done, if we define a D'Alembertian displacement,
$\delta_{i}(t)$ as a variation in $x_{i}(t)$. $dx_{i}$ is a differential
motion over a time $dt$, and is given by $dx_{i}=\dot{x_{i}}dt$.
It vanishes if $dt$ is set to zero. This definition of virtual displacement
is used in \cite{key-16}. A true expression of the constraints on
the variation is,\begin{equation}
a_{i}\dot{\delta_{i}}+a_{ix_{j}}\dot{x_{i}}\delta_{j}+a_{tx_{j}}\delta_{j}=0\label{eq:11}\end{equation}
D'Alembertian displacements don't say anything about the derivatives
of the displacements. They are variations in position, by definition.
As shown earlier, the same constraint force can be obtained by a different
route using Gaussian principle or the Jourdain principle. But the
force there is expressed as $Q_{i}=\lambda(t)\text{\ensuremath{\text{\ensuremath{\mathscr{F}}}}}_{\dot{x_{i}}}$
for non holonomic constraints, where $\ensuremath{\text{\ensuremath{\mathscr{F}}}}=a_{i}\dot{x_{i}}+a_{t}$.
Moreover, Jourdain displacements are virtual velocities, not virtual
displacements. Hence it is impossible to deduce the constraint force
$Q_{i}=\lambda(t)\text{\ensuremath{\text{\ensuremath{\mathscr{F}}}}}_{\dot{x_{i}}}$
directly from D'Alembert's virtual displacement principle, even for
the case of linear non holonomic constraints.

Hence, equation \ref{eq:9}, which is the basis of all support to
\ref{eq:B1}\ref{eq:B2} and all criticism on \ref{eq:C} is incorrect.
It means, the linear non holonomic too, is outside the scope of D'Alembert's
principle, and hence is no way of deciding between \ref{eq:B1}\ref{eq:B2}
and \ref{eq:C}. There is no theoretical way to decide between \ref{eq:B1}\ref{eq:B2}
and \ref{eq:C}. Neither of them are \textit{intrinsically} incorrect.
However, \ref{eq:B1}\ref{eq:B2} seems to be accepted in most practical
examples. \medskip{}

\section*{V. Concluding Remarks}

To summarize, there are two approaches to mechanics. One based on
a constraint force postulate, namely the Jourdain/Gaussian principle(that
covers the D'Alembertian also) and the other based on a variational
principle, namely, the Hamilton's principle. Mechanics based on Hamilton's
principle is called Vakonomic mechanics. More theoretical aspects
of the Vakonomic mechanics can be found in\cite{key-17}. The D'Alembert's
principle is the most intuitive among all. It uses the intuitive concept
of a constraint force acting instantaneously normal to the constraint
surface, and hence forcing the particle to the surface. And hence
it is most reliable. Earlier papers have attempted to compare the
two approaches by checking their consistency with the D'Alembert's
principle. In this paper, we showed that such a comparison is impossible.
However, with an application point of view, the constraint approach
is found to be more oftenly used.\medskip{}

\section*{VI. Appendix: Calculus Of Variation}

A detailed account of calculus of variation can be found any standard
text on the subject\cite{key-18,key-19,key-20} for example). This
section contains a brief outline of calculus of variation in a convenient
format, including all the results relevant for this paper. This whole
section refers to \cite{key-18,key-19}. \medskip{}

Calculus is the study of local variations. For a functional \ensuremath{\mathscr{L}}(f,f',t)
of a function f(t), and its integral defined as 

\begin{equation}
I(f)=\intop_{t_{1}}^{t_{2}}\text{\ensuremath{\mathscr{L}}}(f,f',t)dt\label{eq:12}\end{equation}
where $f'$ is the time derivative of $f$. The local variation in
$I(f)$ for a local variation of $f_{i}(t)$ from $f_{i}(t)$ to $f_{i}(t)+$$\delta_{i}(t)$
is given by\begin{equation}
I(f+\delta)-I(f)=\intop_{t_{1}}^{t_{2}}(\text{\ensuremath{\text{\ensuremath{\mathscr{L}}}}}_{f_{i}}\delta_{i}+\text{\ensuremath{\mathscr{L}}}_{f_{i}'}\delta_{i}')dt+\intop_{t_{1}}^{t_{2}}(\text{\ensuremath{\text{\ensuremath{\mathscr{L}}}}}_{f_{i}f_{j}}\delta_{i}\delta_{j}+\text{\ensuremath{\text{\ensuremath{\mathscr{L}}}}}_{f_{i}f_{j}'}\delta_{i}\delta_{j}'+\text{\ensuremath{\text{\ensuremath{\mathscr{L}}}}}_{f_{i}'f_{j}'}\delta_{i}'\delta_{j}')dt\label{eq:13}\end{equation}
up to the second order. Under the additional condition that $f_{i}(t_{1})$and
$f_{i}(t_{2})$ are fixed during the variation i.e., $\delta_{i}(t_{1})=\delta_{i}(t_{2})=0,$
the first order variation in $I(f)$ can be written as,

\begin{equation}
I(f+\delta)-I(f)=\intop_{t_{1}}^{t_{2}}\{\text{\ensuremath{\mathscr{L}}}_{f_{i}}-\frac{\text{d}}{dt}(\text{\ensuremath{\mathscr{L}}}_{f_{i}'})\}\delta_{i}dt\label{eq:14}\end{equation}
\medskip{}

\subsection*{1.) Unconstrained Extremization}

Extremization of I(f) would mean that finding a f(t) where its first
order variation vanishes. i.e, $\intop_{t_{1}}^{t_{2}}\{\text{\ensuremath{\mathscr{L}}}_{f_{i}}-\frac{\text{d}}{dt}(\text{\ensuremath{\mathscr{L}}}_{f_{i}'})\}\delta_{i}dt=0$for
all functions $\delta_{i}$. That is equivalent to the n differential
equations $\text{\ensuremath{\mathscr{L}}}_{f_{i}}-\frac{\text{d}}{dt}(\text{\ensuremath{\mathscr{L}}}_{f_{i}'})=0$
with the boundary conditions given by $f_{i}(t_{1})$and $f_{i}(t_{2})$\medskip{}

\subsection*{2(a) Constrained Extremization. }

We first consider a constraint of the following type:$\intop_{t_{1}}^{t_{2}}\text{\ensuremath{\mathscr{F}}}(f,f',t)dt=0$
where, $\text{\ensuremath{\mathscr{F}}}(f,f',t)$ is a functional.
This problem is now equivalent finding functions $\{f_{i}(t)\}$ such
that,

(i) They satisfy the constraint $\text{\ensuremath{\text{\ensuremath{\mathscr{F}}}}}(f,f',t)$

(ii)First order variation of I(f) vanishes under local variations
in$\{f_{i}(t)\}$ consistent with the constraint $\text{\ensuremath{\text{\ensuremath{\ensuremath{\text{\ensuremath{\mathscr{F}}}}}}}}(f,f',t)$
and the boundary conditions on f(t)

(iii)$\{f_{i}(t)\}$satisfy the boundary conditions.

Let $\{\delta_{i}(t)\}$be a set of n functions and $\{\epsilon_{i}\}$
be a set of n real numbers. Condition (ii) can be written as:

\begin{equation}
\intop_{t_{1}}^{t_{2}}\{\text{\ensuremath{\mathscr{L}}}_{f_{i}}-\frac{\text{d}}{dt}(\text{\ensuremath{\mathscr{F}}}_{f_{i}'})\}\delta_{i}\epsilon_{i}dt=0\label{eq:15}\end{equation}
for all $\{\epsilon_{i}\}\in S$ where $S=\{\{\epsilon_{i}\}:\epsilon_{i}\intop_{t_{1}}^{t_{2}}\{\text{\ensuremath{\mathscr{F}}}_{f_{i}}-\frac{\text{d}}{dt}(\text{\ensuremath{\mathscr{F}}}_{f_{i}'})\}\delta_{i}dt=0\}$.
For all sets $\{\delta_{i}(t)\}$. Essentially, S is the set of all
variations consistent with the constraints. Such a definition of S
will not work if $\text{\ensuremath{\mathscr{F}}}_{f_{i}}-\frac{\text{d}}{dt}(\text{\ensuremath{\mathscr{F}}}_{f_{i}'})=0$
whenever $\ensuremath{\text{\ensuremath{\mathscr{F}}}}=0$ for all
i. The second order variation should be invoked in such cases. For
a particular set $\{\delta_{i}\}$, let $\alpha_{i}=\intop_{t_{1}}^{t_{2}}\{\text{\ensuremath{\mathscr{L}}}_{f_{i}}-\frac{\text{d}}{dt}(\text{\ensuremath{\mathscr{L}}}_{f_{i}'})\}\delta_{i}dt$
and $\beta_{i}=\intop_{t_{1}}^{t_{2}}\{\text{\ensuremath{\mathscr{F}}}_{f_{i}}-\frac{\text{d}}{dt}(\text{\ensuremath{\mathscr{F}}}_{f_{i}'})\}\delta_{i}dt$.
S is now the set of n-vectors $\{\epsilon_{i}\}$ orthogonal to the
vector $\beta$. Hence the condition says, $\alpha_{i}=\lambda\beta_{i}$
for a real number $\lambda$. A simple argument can establish that
the number $\lambda$ is independent of the set $\{\delta_{i}\}$.
Hence, we have the final equations:

$\alpha_{i}=\lambda\beta_{i}$ for all sets $\{\delta_{i}\}$. Or,
equivalently,

\begin{equation}
\text{(\ensuremath{\mathscr{L}}}-\lambda\text{\ensuremath{\mathscr{F}}})_{f_{i}}-\frac{\text{d}}{dt}(\text{(\ensuremath{\mathscr{L}}}-\lambda\text{\ensuremath{\mathscr{F}}})_{f_{i}'})=0\label{eq:16}\end{equation}
$\lambda$ is the Lagrange multiplier for this case. This is easily
generalized to several constraints. \medskip{}

\subsection*{2(b) Constrained Extremization: constraints of second type.}

Now we consider constraints of the form $\text{\ensuremath{\mathscr{F}}}(f,f',t)=0.$
These constraints can be equivalently written as 

$\intop_{t_{1}}^{t_{2}}k(t)\text{\ensuremath{\mathscr{F}}}(f,f',t)dt=0$
for all functions $k(t)$, or equivalently, for $k(t)=k_{j}(t)$,
where $\{k_{j}(t)\}$ form a basis for functions on $[t_{1},t_{2}]$
Now, the constraint has reduced to the previous category of constraints.
The equations may now be written as:

\begin{equation}
\text{(\ensuremath{\mathscr{L}}}-\lambda(t)\text{\text{\ensuremath{\mathscr{F}}}})_{f_{i}}-\frac{\text{d}}{dt}(\text{(\ensuremath{\mathscr{L}}}-\lambda(t)\text{\text{\ensuremath{\mathscr{F}}}})_{f_{i}'})=0\label{eq:17}\end{equation}
where $\lambda(t)$ is the Lagrange multiplier for this case.

\end{document}